\documentclass[final,3p,times,twocolumn]{elsarticle}
 \biboptions{comma,sort&compress}
\usepackage{here}
 \usepackage{graphicx}
  \usepackage{epsfig}

\def\nin{\noindent}
\def\beq{\begin{equation}}
\def\eeq{\end{equation}}
\def\bea{\begin{eqnarray}}
\def\eea{\end{eqnarray}}

\journal{Nuc. Phys. (Proc. Suppl.)}

\begin{document}

\begin{frontmatter}



\title{Higgs searches at the Tevatron}

 \author[label1]{Paolo Mastrandrea}
 \address[label1]{INFN, Sezione di Roma, 
   Piazzale A. Moro 2, 00185 Roma, Italy}
 \ead{paolo.mastrandrea@roma1.infn.it}
 \author{\\\emph{on behalf of} CDF \emph{and} D\O {} \emph{collaborations}}

\begin{abstract}
\noindent
The search for the Standard Model Higgs boson in $p \bar{p}$ collisions
at 1.96 TeV performed by CDF and D\O{} collaborations at the Tevatron
collider is reported in this paper.
The Higgs candidate events are reconstructed using different final states
in order to optimize the sensitivity in the full range of the Higgs mass.
The presented results use different statistical samples collected
by the Tevatron up to 5.9 fb$^{-1}$.
Combining the most updated limits provided by the two experiments for
all the final states analyzed, the Standard Model Higgs boson
is excluded at 95 \% C.L. in the mass range \mbox{158 - 175 GeV/c$^2$},
in good agreement with the prediction for the analyzed data sample.

\end{abstract}


\end{frontmatter}


\section{Introduction}
\nin

The Standard Model of field and particles (SM)
\cite{Glashow:1961tr}, \cite{Salam:1968rm}, \cite{Weinberg:1976gm}
is the theory
that provides the best description of the known
phenomenology of the particle physics up to now.
It is a quantum field theory based on the
gauge symmetry group
\mbox{$S\!U (3)_{C} \times S\!U (2)_{L} \times U (1)_{Y}$},
with spontaneous symmetry breaking.
This gauge group includes the color symmetry group of the
strong interaction, \mbox{$S\!U (3)_{C}$}, and the symmetry group
of the electroweak interactions,
\mbox{$S\!U (2)_{L} \times U (1)_{Y}$}.
The formulation of the Standard Model as a gauge theory
guarantees its renormalizability, but forbids explicit
mass terms for fermions and gauge bosons.
The masses of the particles are generated in a gauge-invariant
way by the \emph{Higgs Mechanism}
\cite{HiggsMechanism}
via a spontaneous breaking of the electroweak symmetry.
This mechanism also implies the presence of a massive scalar
particle in the mass spectrum of the theory,
the \emph{Higgs boson}.
This particle is the only one, among the basic elements for
the minimal formulation of the SM, to have
not been confirmed by the experiments yet.
In the SM, the mass of the Higgs boson is given by
$m_H = \sqrt{\lambda / 2} \cdot \emph{u}$,
where $\lambda$ is the Higgs self-coupling parameter and
$u$ is the vacuum expectation value of the Higgs field,
\mbox{$u = (\sqrt{2}G_F)^{-1/2} \approx 246$ GeV/c$^2$},
fixed by the Fermi coupling $G_F$, which is determined precisely
from muon decay measurements \cite{GF_measurement}.
Since $\lambda$ is presently unknown, the value of the $m_H$
can not be predicted.
However theoretical arguments \cite{range_arguments} can set approximate
bounds on the range of possible values for the mass of the SM Higgs boson.

The couplings of the SM Higgs boson to fundamental fermions are
proportional to the fermion masses, and the couplings to bosons are
proportional to the squares of the boson masses.
In particular, the SM Higgs boson is a CP-even scalar,
and its couplings to gauge bosons, Higgs bosons and
fermions are given by:
\begin{eqnarray*}
  g_{Hf\overline{f}} = \frac{m_f}{u} , \quad
  g_{HVV}           = \frac{2m^2_V}{u} ,\quad
  g_{HHVV}          = \frac{2m^2_V}{u^2} ,\\
  g_{HHH}           = \frac{3m^2_H}{u} ,\quad\quad
  g_{HHHH}          = \frac{3m^2_H}{u^2}\quad\quad\quad
\end{eqnarray*}
where $V = W^{\pm}$ or $Z$.
In Higgs boson production and decay processes, the dominant
mechanisms involve the coupling of the $H$ to the $W^{\pm}$, $Z$
and/or the third generation quarks and leptons.
The Higgs boson’s coupling to gluons, $Hgg$, is induced
at leading order by a one-loop graph in which the H couples
to a virtual $t\overline{t}$ pair.
Likewise, the Higgs boson’s coupling to photons,
$H\gamma\gamma$, is also generated via loops, although in this
case the one-loop graph with a virtual $W^+ W^-$ pair provides the
dominant contribution \cite{higgs_coupling}.

\begin{figure}[t]
  \centerline{\includegraphics[width=7.cm]{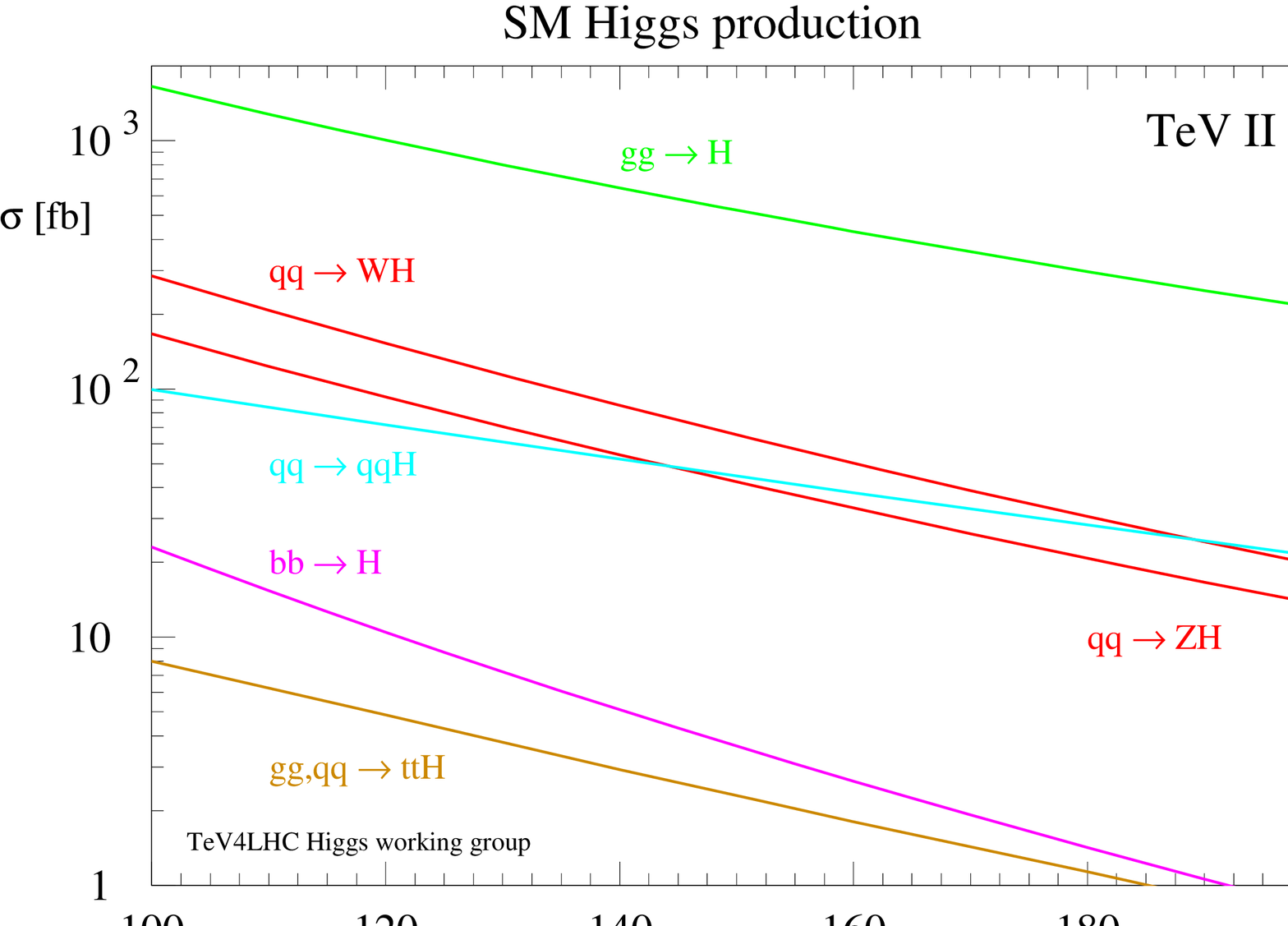}}
  \caption{\scriptsize Standard Model Higgs boson production cross section
    for $p\overline{p}$ collisions at 1.96 TeV .}
  \label{fig1} 
\end{figure}

The cross sections for the production of SM Higgs bosons
are summarized in Fig.\ref{fig1} for $p\overline{p}$
collisions at the Tevatron.

\begin{figure}[t]
  \centerline{\includegraphics[width=7.cm]{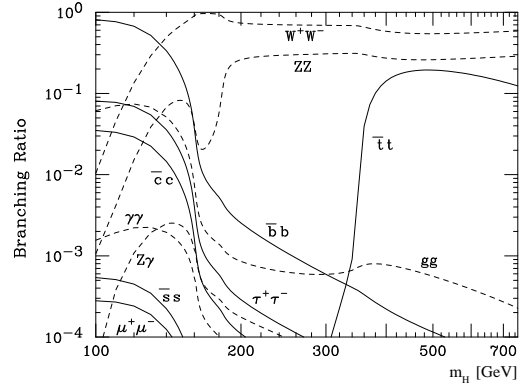}}
  \caption{\scriptsize Branching ratios for the Standard Model Higgs boson.}
  \label{fig2} 
\end{figure}

The branching ratios for the most relevant decay modes of
the SM Higgs boson are shown in Fig.\ref{fig2} as functions of $m_H$.
For masses below 135 GeV/c$^2$ (the so-called \emph{low-mass} region),
decays to fermion pairs dominate, of which the decay $H \to b\overline{b}$
has the largest branching ratio.
Decays to $\tau^+ \tau^-$, $c\overline{c}$ and gluon pairs together
contribute less than 15\%.
For such low masses, the total decay width is less than \mbox{10 MeV}.
For Higgs boson masses above 135 GeV/c$^2$(the so-called \emph{high-mass} region),
the $W^+ W^-$ decay dominates (below the $W^+ W^-$ threshold, one of
the W bosons is virtual) with an important contribution from
$H \to ZZ$, and the decay width rises rapidly, reaching about
1 GeV/c$^2$ at $m_H$ = 200 GeV/c$^2$ and 100 GeV/c$^2$ at $m_H$ = 500 GeV/c$^2$.
Above the $t\overline{t}$ threshold, the branching ratio into top-quark
pairs increases rapidly as a function of the Higgs boson mass,
reaching a maximum of about 20\% at $m_H \sim$ 450 GeV/c$^2$.

In the \emph{low-mass} region the process $gg \to H \to b\overline{b}$
has the biggest cross section, but is not valuable because of the
multijet background, which is overwhelming in an hadronic collider.
The associated production of the Higgs boson and a $W^{\pm}$ or a $Z$
is more rare, but provide a clearer experimental signature.
In the \emph{high-mass} region the process $H \to W^+ W^-$ has the
biggest branching ratio and provide also a clear signature because
of the leptonic decay of the $W$.


\section{Methods}
\nin
All the presented analysis have been performed utilizing the data collected by
the CDF and D\O{} detectors installed at the Fermilab's Tevatron, an
accelerator capable to collide proton with antiproton at a center-of-mass
energy $\sqrt{s} = 1.96$ TeV and an instantaneous luminosity
$\mathcal{L} = 4 \cdot 10^{32}$ cm$^{-2}$ s$^{-1}$.
Both CDF and D\O{} are general purpose detectors, cylindrically symmetric
around the beam axis which is oriented as the $z$ direction.
The polar angle $\theta$ is measured from the origin of the coordinate system
at the center of the detector with respect to the $z$ axis.
The pseudorapidity, transverse energy and transverse momentum are defined as
$\eta = -\ln \tan (\theta / 2)$, $E_T = E \sin (\theta)$ and $p_T = p \sin (\theta)$,
respectively.
Detailed descriptions of the features of the two detectors can be found in
\cite{CDF_DETECTOR} and  \cite{D0_DETECTOR}.

Several combinations of the $W^{\pm}$, $Z$ and Higgs bosons decay channels are utilized to
reconstruct the candidate Higgs events.
High-$p_T$ electrons and muons are widely used in the reconstruction of the
vector bosons candidates because of the clear signatures provided
in an hadronic environment.
Neutrinos are identified evaluating the missing transverse energy ($\not \!\!\! E_T$)
in the event.
The missing $E_T$ is defined by $\not \!\!\! E_T = | \vec{\not \!\!\! E_T} |$, {}
$\vec{\not \!\!\! E_T} = - \sum_i E_T^i \mathbf{\hat{n}_i} $, {}
where $\mathbf{\hat{n}_i} $ is a unit vector perpendicular to the beam axis
and pointing at the $i^{th}$ calorimeter tower.
The index $i$ runs over all the calorimeter towers with an energy deposit
over a minimum threshold.

The jets are reconstructed using a cone algorithm, with a radius
$R = 0.4$ .
The identification of jets originating from a \emph{b}-quark (\emph{b-tagging}) is pursued
using its long lifetime, high mass and leptonic decay.
The main algorithm applied are based on the Reconstruction of a secondary vertex inside the
jet cone (\emph{SecVtx}), on the distribution of the tracks inside the jet cone, respect
to the primary vertex (\emph{JetProbability}), on the reconstruction of a electron or muon
inside the jet cone (\emph{Soft Lepton Tagging}) and on artificial Neural Network.

In order to extract the small signal fraction from the large background in the selected samples,
multivariate techniques are widely used.
Furthermore, in several channels, the analyzed sample is divided in sub-samples in order to
increase the sensitivity.


\section{Results}
\nin
The CDF and D\O{} collaborations are pursuing a direct search
for the SM Higgs boson in the 100 $< m_H <$ 200 GeV/c$^2$
mass range.
In the following the most updated results for the main channels
are presented.
Only peculiar features will be discussed.

\subsection{$ZH \to l^+l^- b\overline{b}$}
\nin
\begin{figure}[t]
\centerline{\includegraphics[width=7.cm]{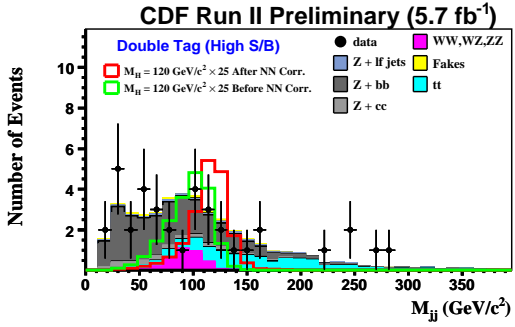}}
\caption{\scriptsize Distribution of the invariant mass of the
selected jets, when both are b-tagged, compared with the expected Higgs
signal, before and after the correction to the jet energy.}
\label{fig21}
\end{figure}
\begin{figure}[b]
\centerline{\includegraphics[width=7.cm]{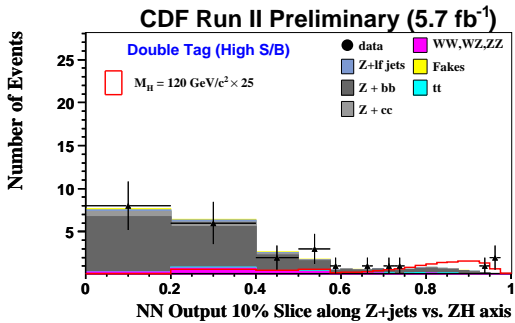}}
\caption{\scriptsize Distribution of the output of the final
Neural Network discriminant for the 2 b-tag category.}
\label{fig22}
\end{figure}
The search for events of associated production of an Higgs
and a $Z$ boson has been performed by CDF collaboration on a data sample
corresponding to 5.7 fb$^{-1}$.
The candidate events are reconstructed using the decays $Z \to l^+l^-$,
where $l = e, \mu$ , and $H \to b\overline{b}$.
The $Z \to l^+l^-$ decays are selected requiring two same-flavor
leptons reconstructing to roughly the Z mass
(76 $\le M_{ll} \le$ 106 GeV/c$^2$).

The events selection requires at least two jets within $|\eta| \le 2$:
the first with $E_T \ge 25$ GeV and the second with $E_T \ge 15$ GeV.

The selected events are divided into six categories
according to the quality of the leptons used to reconstruct the $Z$ candidate
and to the b-tagging of the jets.

The dijet mass ($M_{jj}$) is one of the most useful distributions to separate
the $ZH$ events the $Z+jets$ background, with its separating power
limited mainly by the jet-energy resolution.
A Neural Network based technique is used to correct the jet energies, in
order to increase the $ZH$ vs $Z+jets$ discrimination.

A final Neural Network based discriminant collects the most powerful
kinematic and tagging variable and the Matrix Element probabilities
for each event, providing a limit on the $ZH$ production for each category.
For a SM Higgs boson mass of 120 GeV/c$^2$, the expected 95\% C.L. upper limit
is evaluated to be 6.91 times the SM prediction with an observed limit of 8.32 .

%

\subsection{$ZH \to \nu \nu b\overline{b} \ $ and $ \ WH \to (l)\nu b\overline{b}$}
\nin
\begin{figure}[t]
  \centerline{\includegraphics[width=7.cm]{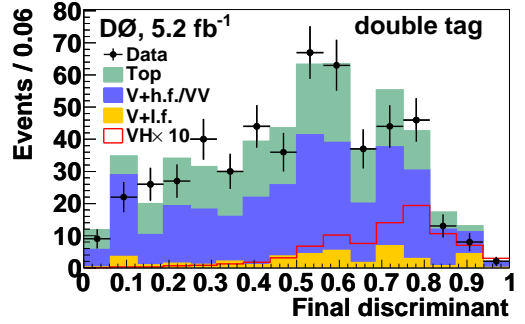}}
  \caption{\scriptsize Distribution of the output of the final
    Neural Network discriminant for the 2 b-tag category for the
    analysis performed by the D\O{} collaboration for the
    $\ \not \!\!\! E_T \ b\overline{b}$ final state.}
  \label{fig23}
\end{figure}

The search for events of low-mass Higgs boson production
in the $\ \not \!\!\! E_T \ b\overline{b}$ final state,
has been performed by CDF and D\O{} collaboration on a data sample
corresponding to 5.7 fb$^{-1}$ and 5.2 fb$^{-1}$ respectively,
with similar techniques.
The candidate events are selected requiring large
$\ \not \!\!\! E_T$ and at least two jets in the region
of high efficiency for the b-tagging algorithms.
An artificial Neural Network is trained to reduce the
multijet background.
A final Neural Network based discriminant is used to evaluate the
limit on the Higgs boson production.

For a SM Higgs boson mass of 115 GeV/c$^2$, the 95\% C.L. upper limit
is observed to be 2.3 times the SM prediction (expected 4.0) by the CDF collaboration,
and is observed to be 3.7 times the SM prediction (expected 4.6) by the D\O{} collaboration.

\subsection{$H \to WW^*$}
\nin
\begin{figure}[t]
  \centerline{\includegraphics[width=7.cm]{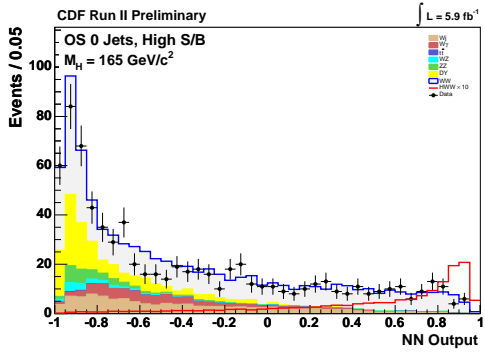}}
  \caption{\scriptsize Distribution of the output of the final
    Neural Network discriminant for the 2 b-tag category for the
    analysis performed by the CDF collaboration.}
  \label{fig33}
\end{figure}
\begin{figure}[!b]
  \centerline{\includegraphics[width=6.cm]{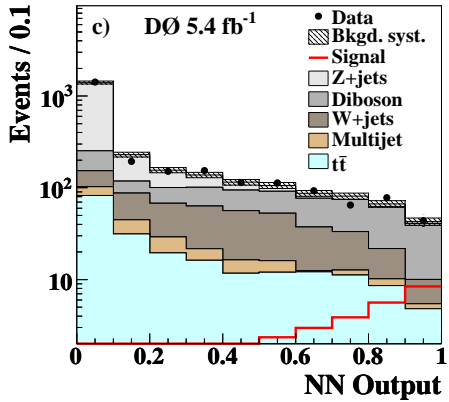}}
  \caption{\scriptsize Distribution of the output of the final
    Neural Network discriminant for the analysis performed by the
    D\O{} collaboration.
    In this case the results for the different categories are
    combined together in a single discriminant.}
  \label{fig34}
\end{figure}
The search for events of high-mass Higgs boson production
has been performed by CDF and D\O{} collaboration on a data sample
corresponding to 5.9 fb$^{-1}$ and 5.4 fb$^{-1}$ respectively,
using similar techniques.
The candidate events are reconstructed using the decays $W \to l \nu$,
where $l = e, \mu$.

The selected events are divided in several categories according to
same/opposite sign of leptons, jet multiplicity, selected leptons
invariant mass ($M_{ll}$).
Same sign and tri-lepton categories provide acceptance also for the
$WH \to WWW^*$ and $ZH \to ZWW^*$ channels.

A set of Neural Network based discriminants is used to separate signal
from background in the different categories.

For a SM Higgs boson mass of 165 GeV/c$^2$, the 95\% C.L. upper limit
is observed to be 1.08 times the SM prediction (expected 1.0) by the CDF collaboration,
and is observed to be 1.55 times the SM prediction (expected 1.36) by the D\O{} collaboration.

\begin{figure}[t]
  \centerline{\includegraphics[width=8.cm]{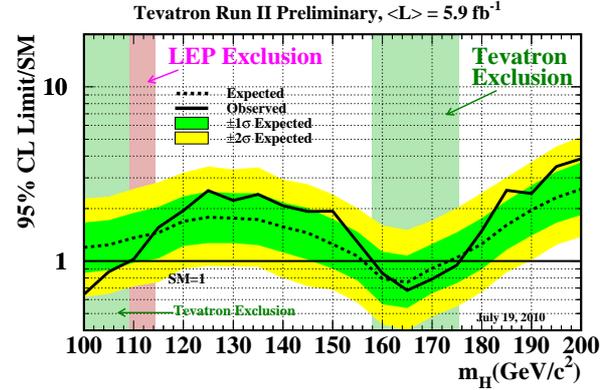}}
  \caption{\scriptsize Ratio of the observed 95 \% C.L. upper limit to the
    SM prediction as a function of the Higgs boson mass.}
  \label{fig_combination}
\end{figure}


\section{Conclusions}
\nin
All the most updated results for each channel from both the experiments
are combined in order to maximize the sensitivity.
The statistics and systematics uncertainties are taken into account in the
combination procedure, and their correlations are treated consistently.
The ratio of the observed 95 \% C.L. upper limit to the SM prediction
is reported as a function of the Higgs mass in Fig. \ref{fig_combination}.
Tevatron experiments exclude SM Higgs boson at 95 \% C.L. for mass
between 158 and 175 GeV/c$^2$, and between 100 and 109 GeV/c$^2$.


\end{document}